\pdfoutput=1
\documentclass[prb,twocolumn,showpacs,amsmath,longtable,amssymb,floatfix,superscriptaddress]{revtex4}

\usepackage{graphicx}%Include figure files
\usepackage{dcolumn}%Align table columns on decimal point
\usepackage{bm}% bold math
\usepackage[T1]{fontenc}
\usepackage{amsmath}
\usepackage{amssymb}
\usepackage{xcolor}
\newcommand*{\Tr}{%
	\mathrm{Tr} }

\begin{document}
	
	\title{Damping and Anti-Damping Phenomena in Metallic Antiferromagnets: An ab-initio Study}
	
	\author{Farzad Mahfouzi}
	\email{Farzad.Mahfouzi@gmail.com}
	\affiliation{Department of Physics and Astronomy, California State University, Northridge, CA, USA}
	\author{Nicholas Kioussis}
	\email{Nick.Kioussis@csun.edu }
	\affiliation{Department of Physics and Astronomy, California State University, Northridge, CA, USA}
	
	\begin{abstract}
We report on a first principles study  of anti-ferromagnetic resonance (AFMR) phenomena in metallic systems [MnX (X=Ir,Pt,Pd,Rh)  and FeRh] under an external electric field. 
We demonstrate that the AFMR linewidth can be separated into a relativistic component originating from the angular momentum transfer between the collinear AFM subsystem and the crystal through the spin orbit coupling (SOC), 
and an exchange component that originates from the spin exchange between the two sublattices. The calculations reveal that the latter component becomes significant in the low temperature regime. Furthermore, we present results for the current-induced intersublattice torque which can be separated into the Field-Like (FL) and Damping-Like (DL) components, affecting the intersublattice exchange coupling and AFMR linewidth, respectively.

	\end{abstract}

	\maketitle
	
%\section{Introduction}\label{sec:intro}
Spintronics is a field of research exploiting the mutual influence between the electrical field/current and the magnetic ordering.
Todate the realization of conventional spintronic devices has relied primarily on the ferromagnetic (FM) based heterostructures\cite{Slonczewski1996,Berger1996,Manchon2008,Miron2010,Miron2011,Liu2012}. On the other hand, antiferromagnetic  (AFM) materials, have been recently revisited as potential alternative candidates for active elements in spintronic devices\cite{Baltz2018,Gomonay2014}. In contrast to their FM counterparts, AFM systems have weak sensitivity to magnetic field perturbations, produce no perturbing stray fields, and can offer ultra-fast writing schemes in terahertz (THz) frequency range. The THz spin dynamics due to AFM ordering has been experimentally demonstrated using all-optical\cite{Kirilyuk2010,Wienholdt2012}, and N\'{e}el SOT\cite{Wadley2016,Bhattacharjee2018} mechanisms.

%%%%%%%%%%%%%%%%%%%%%%  DAMPING FOR FMs and AFMs     %%%%%%%%%%%%%%%%%%%%%%%
One of the most important parameters in describing the dynamics of the magnetic materials is the Gilbert damping constant, $\alpha$. Intrinsic damping in metallic bulk FMs\cite{Kambersky2007,mahfouziPRB2017_GD} is associated with the coupling between the conduction electrons and the 
time-dependent magnetization, $\vec{m}(t)$, where the latter in the presence of spin-orbit coupling (SOC) leads to a modulation (breathing) of 
 the Fermi surface\cite{Kambersky2007} and hence excitation of electrons near the Fermi energy. The excited conduction electrons in turn relax to the ground state through interactions with the environment ({\it e.g.} phonons, photons, etc), leading to a net loss of the energy/angular momentum in the system. While the damping in FMs has been extensively studied both experimentally and theoretically, the damping in metallic AFM has not received much attention thus far.

%%%%%%%%%%%%%%%%%%%%%%%%%%%%%%%%%%%%    ANTI-DAMPING FOR FM AND AFM %%%%%%%%%%%%%%%%%%%%%%%%%%%%%%%%
Manipulation of the damping constant in magnetic devices is one of the prime focuses in the field of spintronics. Conventional approaches to manipulate the damping rate of a FM rely on the injection of a spin polarized current into the FM. The spin current is often generated either through the Spin Hall Effect (SHE)\cite{Dykanov1971,Sinova2015} by a charge current passing through a heavy metal (HM) adjacent to the FM in a lateral structure, or spin filtering in a magnetic tunnel junction (MTJ) in a vertical heterostructure\cite{Ralph2008}.
Similar mechanisms have also been proposed\cite{Nunez2006,Gomonay2014,Gulyaev2014,Cheng2016,Gulbrandsen2018,Khymyn2017} for AFM materials, where the goal is often to cause spontaneous THz-frequency oscillations or reorientation\cite{Wadley2016,Kriegner2016,Chen2018,Moriyama12018} of the AFM N\'{e}el ordering, $\vec{n}(t)=(\vec{m}_1-\vec{m}_2)/2$. Here, $\vec{m}_s$ is a unit vector along the magnetization orientation of the sublattice $s$. In contrast to the aforementioned studies that require breaking of inversion symmetry to induce N\'{e}el ordering switching, in this work we focus on the current-induced excitation of the sublattice spin dynamics of bulk metallic AFM materials with inversion symmetry intact, and hence no N\'{e}el SOT\cite{Wadley2016,Bhattacharjee2018,Zelezny2014,Zelezny2017}.

\begin{figure}
		\includegraphics[scale=0.5,angle=0]{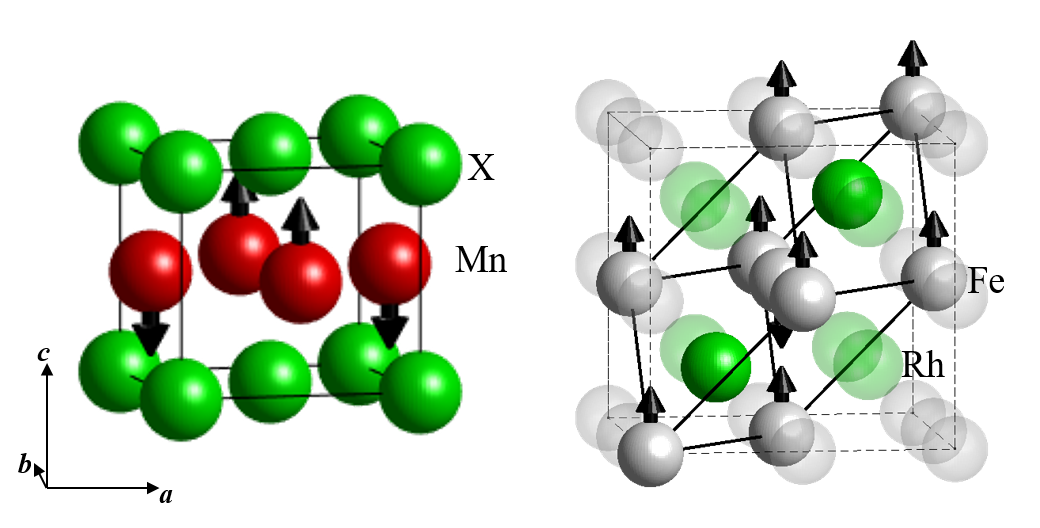}
		\caption{(Color online) Crystal structure of (left:) MnX with (X=Ir,Pt,Pd,Rh) and (right:) FeRh used for the first principle calculations, where the corresponding spin configuration and  primitive cells are shown with solid lines.}
		\label{fig:fig1}
	\end{figure}

%%%%%%%%%%%%%%%%%%%%%%%%%%%%%%%%   OBJECTIVE OF THIS MANUSCRIPT
The objective of this work is to, (1) provide a general analytical expression for the AFMR\cite{Kittel1951} frequency and linewidth in the presence of current-induced sublattice torque, and (2) employ the Kubo-like formalism with first principles calculations to calculate 
the Gilbert damping tensor, $\alpha_{ss'}$ ($s,s'=\uparrow,\downarrow$), and the field-, $\vec{\tau}_{FL}$, and damping-like, $\vec{\tau}_{DL}$, components of the sublattice torque for a family of metallic AFM materials including MnX (X=Ir,Pt,Pd,Rh) and FeRh, shown in Fig.~\ref{fig:fig1}. We demonstrate that the zero-bias AFMR linewidth can be separated into the relativistic,  $\Gamma^{r}=\lambda\alpha_{0}/2M$,  and exchange, $\Gamma^{ex}=\mathcal{K}\alpha_{d}/2M$ components\cite{Mentink2012}, where $\alpha_d\equiv\sum_{s}\alpha_{ss}$, $\alpha_{0}\equiv\alpha_d-\sum_{s}\alpha_{s\bar{s}}$, $M$ is the magnetic moment of each sublattice, $\lambda$ is the intersublattice exchange interaction, and $\mathcal{K}$ is the magnetocrystalline anisotropy energy. In agreement with recent first principles calculations\cite{Liu2017}, we find that $\alpha_d$ is about 3 orders of magnitude larger than $\alpha_0$, indicating the crucial role of the exchange component to the AMFR linewidth. Our calculations reveal that at high temperatures the interband contribution to the  relativistic component is the dominant term in the AFMR linewidth, while at low temperatures both exchange and relativistic components contribute to the AFMR linewidth on an equal footing. 
%%%%%%%%%%%%%%%%%%%%%%  Current-induced modulation of AFMR linewidth and 
We further demonstrate that the current-induced antidamping- (field-) like torque changes the AFMR linewidth (intersublattice exchange interaction), thereby allowing the manipulation of the damping constant (N\'{e}el temperature) in bulk AFM materials.
	
	Precessional magnetization dynamics of AFMs is often described by a system of coupled equations for each spin sublattice,\cite{Cheng2016,Khymyn2017,Checinski2017} where a local damping constant $\alpha$ is assigned to each of the two sublattices ignoring the effects of the rapid (atomic scale) spatial variation of the magnetization on the damping constant due to the AFM ordering. 
	Taking into account the Gilbert damping tensor, $\alpha_{ss'}$, the coupled LLG equations of motion for the two sublattices can be written as,
	%	\begin{subequations}
	%	\end{subequations}
	
	\begin{align}\label{eq:LLG}
	\frac{d\vec{m}_{s}(t)}{dt}=&-\gamma\vec{m}_s(t)\times\vec{H}_s^{eff}+\sum_{s'}\alpha_{ss'}\vec{m}_s(t)\times\frac{d\vec{m}_{s'}(t)}{dt},
	\end{align}	
	where 
	%$\vec{m}_s$ is a unit vector pointing along the direction of the magnetic moment of sublattice $s$, and 
	the local effective field in the presence of the external electric ($\vec{E}_{ext}$) and magnetic ($\vec{B}_{ext}$) fields, is given by,
	\begin{align}\label{eq:BField}
	&\vec{H}_s^{eff}=\vec{B}_{ext}+\sum_{i=xyz}\left(K^{(2)}_{i;s}+K^{(4)}_{s}(1-{m}_{i;s}^2(t))\right)\frac{{m}_{i;s}(t)}{M_s}\hat{e}_i\nonumber\\
	&+e\vec{\tau}_{DL}^{0}\cdot\vec{E}_{ext}\vec{m}_s(t)\times\vec{m}_{\bar{s}}(t)+\Big(\frac{\lambda}{M_s}+e\vec{\tau}_{FL}^{0}\cdot\vec{E}_{ext}\Big)\vec{m}_{\bar{s}}(t).
	\end{align}	
	Here, $\lambda$ is the exchange coupling between the two sublattices, $\vec{\tau}_{DL}^{0}$($\vec{\tau}_{FL}^{0}$) is the current-induced intersublattice damping-like (field-like) torque component and $K^{(2)}_{i;s}$ ($K^{(4)}_{s}$) is the second (fourth) order magneto-crystalline anisotropy energy (MCAE). 
%	Here we did not include the  N\'{e}el SOT\cite{Zelezny2014,Wadley2016,Bhattacharjee2018,Zelezny2017} term
%	since the focus of this work is the sublattice dynamics in materials where 
%	the local inversion symmetry\cite{Zelezny2017} is not broken. 
	Eq.~\eqref{eq:BField} shows that the effect of $\vec{\tau}_{FL}^{0}$ is to renormalize the intersublattice exchange coupling,  $\lambda'=\lambda+M_se\vec{\tau}_{FL}^{0}\cdot\vec{E}_{ext}$. 
	
	In the following, without the loss of generality, we assume $K^z_{2}=0$ and $K^{x,y}_{2}\ge0$, where in the absence of an external magnetic field the magnetization relaxes towards the $\hat{e}_z$-axis which can be either in- or out-of-plane. Consequently, we consider 
	 $\vec{m}_s(t)=m^{z}_s\hat{e}_z+\delta\vec{m}_s(t)$, where, $m^{z}_s=\pm 1$ and $\delta\vec{m}_s(t)$ is small deviation of the magnetic moment normal to the easy ($\hat{e}_z$) axis. Solving the resulting linearized LLG equations of motions, the poles of the transverse dynamical susceptibility yield two oscillating modes with resonance frequencies, $\omega_j$, given by
	\begin{subequations}
	\begin{align}\label{eq:EigenFreqs}
	&(\frac{\omega_j}{\gamma}-i\vec{\tau}^0_{DL}\cdot\vec{E}_{ext})^2=(\omega^0_j)^2+2i\Gamma_{j}\frac{\omega_j}{\gamma},\ j=x,y \\
		&\omega^0_j=\frac{\sqrt{\mathcal{K}_x\mathcal{K}_y+2\lambda' \mathcal{K}_{j}}}{M}
	\end{align}
	\end{subequations}
	where, $M=|M_s|$, $\mathcal{K}_{j}=K_{j}^{(2)}+K^{(4)}$ and the AFMR linewidth  
	\begin{equation}
	 \Gamma_{j} \equiv \Gamma^{r} + \Gamma_j^{ex} = \frac{1}{2M}\big(\lambda'\alpha_{0}+\mathcal{K}_{j}\alpha_d\big),
	\end{equation}
	can be separated into a relativistic component originating from the angular momentum transfer between the collinear AFM orientation and the crystal through the SOC, 
	and an exchange component that originates from the spin current exchange between the two AFM sublattices.
	For a system with uniaxial MCAE, Eq.\eqref{eq:EigenFreqs} can be used in both cases of out-of- and in-plane precessions with $K_{x,y}^{(2)}=|K_{\perp}^{(2)}|$ and $K_{y}^{(2)}=0, K_{x}^{(2)}=|K_{\perp}^{(2)}|$, respectively, where $|K_{\perp}^{(2)}|$ is the amplitude of the out of plane MCAE. Eq.~(\ref{eq:EigenFreqs}) is the central result of this paper which is used to calculate the AFMR frequency and linewidth and their corresponding current-induced effects. A more general form of Eq.~(\ref{eq:EigenFreqs}) in the presence of an external magnetic field along the precession axis is presented in the Appendix.~\ref{app:C}.
	
	Eq.~(\ref{eq:EigenFreqs}) also yields the effective Gilbert damping 
	\begin{align}\label{eq:EffGD}
	\alpha^{eff}_j \equiv \frac{\delta Im(\omega_j)}{\delta Re(\omega_j)} = 
	\frac{\lambda\alpha_{0}+\mathcal{K}_j\alpha_d}{2M\sqrt{\mathcal{K}_x\mathcal{K}_{y}+2\lambda \mathcal{K}_{j}}},\ j=x,y.
	\end{align}
	
\noindent Similarly to the linewidth, $\alpha^{eff}_j$ can be separated into the  relativistic, $\alpha^r_j = \Gamma^r_j/\omega_j^0$ and exchange,  $\alpha^{ex}_j = \Gamma^{ex}_j/\omega_j^0$, contributions.
To understand the origin of the relativistic component of the AFMR linewidth, one can use a unitary transformation into the rotating frame of the AFM direction, where $\alpha_{0}$ can be written in terms of the matrix elements of $\hat{H}_{SOC}$ using the spin-orbital torque correlation (SOTC) expression,\cite{mahfouziPRB2017_GD} also often referred to as Kambersky's formula\cite{Kambersky2007},
\begin{equation}
\alpha_{0}=\frac{\hbar}{\pi N_kM}\sum_{\vec{k}}\Tr(\hat{A}_{\vec{k}}[\hat{H}_{SOC},\sigma^{+}]\hat{A}_{\vec{k}}[\hat{H}_{SOC},\sigma^{-}]).
\end{equation}
Here, $\hat{A}_{\vec{k}}=Im(G_{\vec{k}}^r)$ is the spectral function, $\hat{G}^r_{\vec{k}}$ is the retarded Green function calculated at the Fermi energy,  $2\sigma^{\pm} = \sigma_x \pm i \sigma_y$ are the spin ladder operators, and $N_k$ is the number of k-point sampling in the first Brillouin zone. 

A similar approach applied to the intersublattice elements of the damping tensor leads to a relationship between different elements of $\alpha_{ss'}$, rather than an explicit expression for each element. This is due to the fact that in the rotating frame of one sublattice, the other sublattices precesses. Therefore, to calculate  $\alpha_d$ we employ the original torque correlation expression\cite{mahfouziPRB2017_GD} ,
	\begin{align}\label{eq:Gilbert_SOTC}
		\alpha_{d}&=\sum_s\frac{\hbar}{\pi N_kM}\sum_{\vec{k}}\Tr(\hat{A}_{\vec{k}}\hat{\Delta}_{\vec{k}}^s\hat{\sigma}^+\hat{A}_{\vec{k}}\hat{\Delta}_{\vec{k}}^{s}\hat{\sigma}^-)
		\end{align}	
where $\hat{\Delta}_{\vec{k}}^s$ is the exchange spitting of the conduction electrons for sublattice $s$.

	Since, for AFMs with N\'{e}el temperature above room temperature $\lambda\gg \mathcal{K}_j$, one might conclude that $\alpha^{r}\gg \alpha^{ex}$ and the effects of the intersublattice spin exchange on the AFMR line-width becomes negligible. However, since $|\alpha_{ss}|$ is proportional to the intersublattice hopping strength[see Appendix.~\ref{app:B}] one can expect to have $\|\alpha_{ss'}\|\gg\alpha_{0}$. 
Therefore, the interplay between the  relativistic and exchange terms is material dependent, where, for systems with $\lambda\gg \mathcal{K}_j$,  the effect of the intersublattice spin exchange on the AFMR linewidth may dominate.

%	\begin{widetext}		
%		\begin{table}
%			\begin{tabular}{|| c c c c c c c||} 
%				\hline
%				& FeRh & MnRh & MnPd & MnPt & MnIr & Mn$_2$Au\\ 
%				[0.5ex] 
%				\hline\hline
%				$10^3\alpha^r_{\perp}$       $\begin{array}{c} \eta\\\frac{\eta}{10}\end{array}$& $\begin{array}{c} \frac{1.7}{\sqrt{x}}\\\frac{0.6}{\sqrt{x}}\end{array}$  &$\begin{array}{c}10\\5\end{array}$&  $\begin{array}{c} 1.6\\1.3\end{array}$ & $\begin{array}{c} 6.7\\4.6\end{array}$   &$\begin{array}{c} 39\\12\end{array}$    & $\begin{array}{c}49\\448 \end{array}$    \\
%				\hline
%				$10^3\alpha^{ex}_{\perp}$       $\begin{array}{c} \eta\\\frac{\eta}{10}\end{array}$& $\begin{array}{c} 1.5\sqrt{x}\\15\sqrt{x}\end{array}$  &$\begin{array}{c}0.6\\2\end{array}$&  $\begin{array}{c} 0.9\\5.6\end{array}$ & $\begin{array}{c} 1.2\\10\end{array}$   &$\begin{array}{c} 3.6\\6\end{array}$    & $\begin{array}{c}2.9\\27 \end{array}$    \\
%				\hline
%			\end{tabular} 
%			\label{table:tab2}
%		\end{table}
%	\end{widetext}

%\section{Results and Discussion}\label{sec:results}

 \begin{table*} [t]
 	\caption{ Calculated sublattice magnetic moment ($M_s$), magnetocrystalline anisotropy energy 
 		per unit cell, ($K_2^{\perp}$), intersublattice exchange coupling per unit cell ($\lambda$), 
 		ratio of the resistivity ($\rho_{xx}$) to the broadening parameter, $\eta$, and the experimental values of $\rho_{xx}$.  We also list 
 		values of $\alpha_d$, $\alpha_0^{\vec{m}_s\parallel \vec {a}(\vec{c})}$ for sublattice 
 		magnetization parallel to the $\vec{a}$($\vec c$) axis, the relativistic  ($\alpha^r_{\perp}$) and exchange ($\alpha^{ex}_{\perp}$) damping parameter for out of plane oscillation mode, for $\eta$ and $\eta$/10 corresponding 
 		to the high- and low-temperature regimes, respectively. Finally, we list values of  
 		the sublattice current-induced field-like $\big({\tau}_{FL}^{0,{\vec{E}\parallel \vec{a}(\vec{c})}}\big)$ and antidamping-like (${\tau}_{DL}^{0,{\vec{E}\parallel \vec{a}(\vec{c})}}$) components of the spin-orbit torques under an external electric field 
 		along the $\vec{a}$ ($\vec{c}$)-axis for room-temperature broadening. }	
 	\begin{ruledtabular}
 		\begin{tabular}{cccccccccccccccc}
 			&  $|M_s|$  &  $c/a$  & $K_2^{\perp}$ & $\lambda$
 			& 	$\rho_{xx}/\eta$ 
 			& $\rho^{exp}_{xx}$
 			& $\eta$
 			& $\alpha_d$ 
 			& $\alpha_0^{\vec{m}_s\parallel \vec{c}}$
 			& $\alpha_0^{\vec{m}_s\parallel \vec{a}}$
 			& $\alpha^r_{\perp}$ 
 			& $\alpha^{ex}_{\perp}$ 
 			& ${\tau}_{FL}^{0,{\vec{E}\parallel \vec{a}(\vec{c})}}$
 			& ${\tau}_{DL}^{0,{\vec{E}\parallel \vec{a}(\vec{c})}}$
 			\\
 			&  ($\mu_B$) &          &  (meV)  & (eV)
 			& $(\frac{\mu\Omega cm}{meV})$
 			& $(\mu\Omega cm)$
 			& (meV)
 			&
 			& ($10^{-3}$) &  ($10^{-3}$) & ($10^{-3}$) & ($10^{-3}$)
 			& ($10^{-3}$\AA) & ($10^{-3}$\AA)
 			\\
 			
 			%%%%%%%%%%%%%%%%%%%%%%%%%%%%%%%%%%%%%%%%%%%%%%%%%%%%%%%%%%%%%%%%%%%%%%%%%%%%%%%%%%%%%%%
 			\hline
 			%%%%%%%%%%%%%%%%%%%%%%%%%%%%%%%%%%%%%%%%%%%%%%%%%%%%%%%%%%%%%%%%%%%%%%%%%%%%%%%%%%%%%%%%%%%
 			FeRh
 			%%%%%%%%%%%%%%%%%%%%%%%%%%%%%%%%%%%%%%%%%%%%%%%%%%%%%%%%%%%%%%%%%%%
 			& 3.1  & 1+$x$ & -1.2$x$  & 0.44  &  3.4 
 			& $\approx$ 100 \footnotemark[1]        & 29    
 			& 0.25  &  0.8  & 0.8 
 			&  $1.7/\sqrt{|x|}$    &   1.5$\sqrt{|x|}$    &  33 (33)   &  -14 (-14) 
 			\\
 			%%%%%%%%%%%%%%%%%%%%%%%%%%%%%%%%%%%%%%%%%%%%%
 			& & & & & &  & 2.9 
 			& 2.5  & 0.27 & 0.27 & $0.6/\sqrt{|x|}$
 			& 15$\sqrt{|x|}$ & & 
 			\\
 			%%%%%%%%%%%%%%%%%%%%%%%%%%%%%%%%%%%%%%%%%%%%%%%%%%%%%%%%%%%%%%%%
 			\\
 			MnRh  &  3.1 & 0.94  &  -0.7 & 0.42 
 			&  0.57 &   95\footnotemark[2]     &  166  &  0.13   
 			&  3.3  & 3.9 & 10 & 0.6 & 10 (7)  & 6 (-3) 
 			\\
 			& & & & & &  & 16.6 
 			& 0.45  & 1.5 & 1.7 & 5
 			& 2  &  & 
 			\\
 			%%%%%%%%%%%%%%%%%%%%%%%%%%%%%%%%%%%%%%%%%%%%%%%%%%%%%%%%%%%%%%%
 			MnPd  
 			&  3.9 & 0.93  &  -0.6 & 0.5 
 			&  2.6 &   223 \footnotemark[3]     &  103  &  0.3   
 			&  0.5  & 0.6  & 1.6  & 0.9  & -2 (-5)  & 93 (4)
 			\\
 			& & & & & &  & 10.3 
 			& 1.8  & 0.1 & 0.5  &  1.3
 			& 5.6  &  &  
 			\\
 			%%%%%%%%%%%%%%%%%%%%%%%%%%%%%%%%%%%%%%%%%%%%%%%%%%%%%%
 			MnPt 
 			&  3.8 & 0.93  & 0.45  & 0.48 
 			&  2.7 &   119,\footnotemark[4]164\footnotemark[5]     &  48  &  0.43   
 			&  2.2  & 7.1  & 6.7  & 1.2  & -15 (17)  & 1 (11)
 			\\
 			& & & & & &  & 4.8 
 			& 3.5  & 1.5 & 21  &  4.6
 			& 10  &  &  
 			\\
 			%%%%%%%%%%%%%%%%%%%%%%%%%%%%%%%%%%%%%%%%%%%%%%%%%%%%%%%%%%%%%%
 			MnIr
 			&  2.6 & 0.97  & -5.9  & 0.4 
 			&  0.5  &   176-269\footnotemark[6]     &  350  &  0.22  
 			&  36  & 35  & 39  & 3.6  & 7 (13)  & 18 (-7)
 			\\
 			& & & & & &  & 35
 			& 0.36  & 14 & 11  &  12
 			& 6  &  &   
 			\\
 			%%%%%%%%%%%%%%%%%%%%%%%%%%%%%%%%%%%%%%%%%%%%%%%%%%%%%%%%%%%%%%%%	
 		\end{tabular}
 	\end{ruledtabular}\label{table:tab1}
 	\footnotetext[1]{Ref.\cite{Mankovsky2017}; ~ $^b$Ref.\cite{KOUVEL1963}; 
 ~	$^c$Ref.\cite{Zhang2014}; ~ $^d$Ref.\cite{Ou2016}; ~$^e$Ref.\cite{Kim2016,Moriyama2014,Tshitoyan2015,Zhang2014}}	

 \end{table*}
 
 %%%%%%%%%%%%%%%%%%%%%%%%%%%%%%%%%%%%%%%%%%   RESULTS %%%%%%%%%%%%%%%%%%%%%%%%%%%%%%%%%%%%%%%%%%%
 %%%%%%%%%%%%%%%%%%%%%%%%%%%%%%%%%%%%%%%%%%%%%%%%%%%%%%%%%%%%%%%%%%%%%%%%%%%%%%%%%%%%%%%%%%%%
 	The crystal structure, conventional and primitive cell, and the AFM ordering of the MnX (X=Pt,Pd,Ir,Rh) family of metallic bulk AFMs and the biaxially strained AFM bulk FeRh 
 	is shown  in  Fig.~\ref{fig:fig1}. The details of the electronic structure calculations of the various	damping and antidamping properties are described in detail in the Appendix.~\ref{app:DFT_calc}.
 	Table~\ref{table:tab1} lists the {\it ab initio} results of the sublattice magnetic moment, $M_s$,
 	$c/a$ ratio, magnetocrystalline anisotropy energy, $K_{\perp}^{(2)}$ , intersublattice exchange
 	 interaction, $\lambda$, and ratio of the longitudinal conductivity to the broadening parameter, $\rho_{xx}/\eta$, for the FeRh and MnX systems, respectively. We also list experimental values of 
 the room-temperature $\rho_{xx}$ which were used to determine the broadening parameter. For FeRh we provide the linear
 	dependence of  $K_{\perp}^{(2)}$ as a function of biaxial strain, $x\equiv c/a -1$, which shows 
 	that under compressive (tensile) biaxial strain the magnetization is along the $c$ ($a$) axis\cite{Bordel2012}. For the MnX family the magnetization is along the $a$ axis except for MnPt.
 	The MCA values for both MnX and FeRh are in good agreement with previous {\it ab-nitio} calculations \cite{Umetsu2006,Bordel2012,Shick2010,Chang2018}. 
 	
 	We also list in	Table~\ref{table:tab1} values of $\alpha_d$ and $\alpha_0^{\vec{m}_s\parallel \vec {a}(\vec{c})}$ for sublattice  magnetization parallel to the $\vec{a}$($\vec c$) axis, and the 
 	relativistic  ($\alpha^r_{\perp}$) and exchange ($\alpha^{ex}_{\perp}$) damping components 
 	of the effective Gilbert damping  for $\eta$ at room temperature and 
 	$\eta/10$ corresponding to low temperature. 
 	The decrease (increase) of the damping constants with decreasing temperature is associated with the conductivity (resistivity)-like regime where the inter-(intra-) band scattering contribution is dominant.  
 	We find that for the $\eta$ value corresponding to room temperature the AFMR linewidth is mostly dominated by the relativistic component, while at low temperatures the two components are comparable in magnitude. For FeRh a relatively large strain (i.e. $x\approx0.1$)  is required
 	to render the exchange component have a significant contribution to the AFMR linewidth 
 	at low temperature.

\begin{figure}
	\includegraphics[scale=0.23,angle=0]{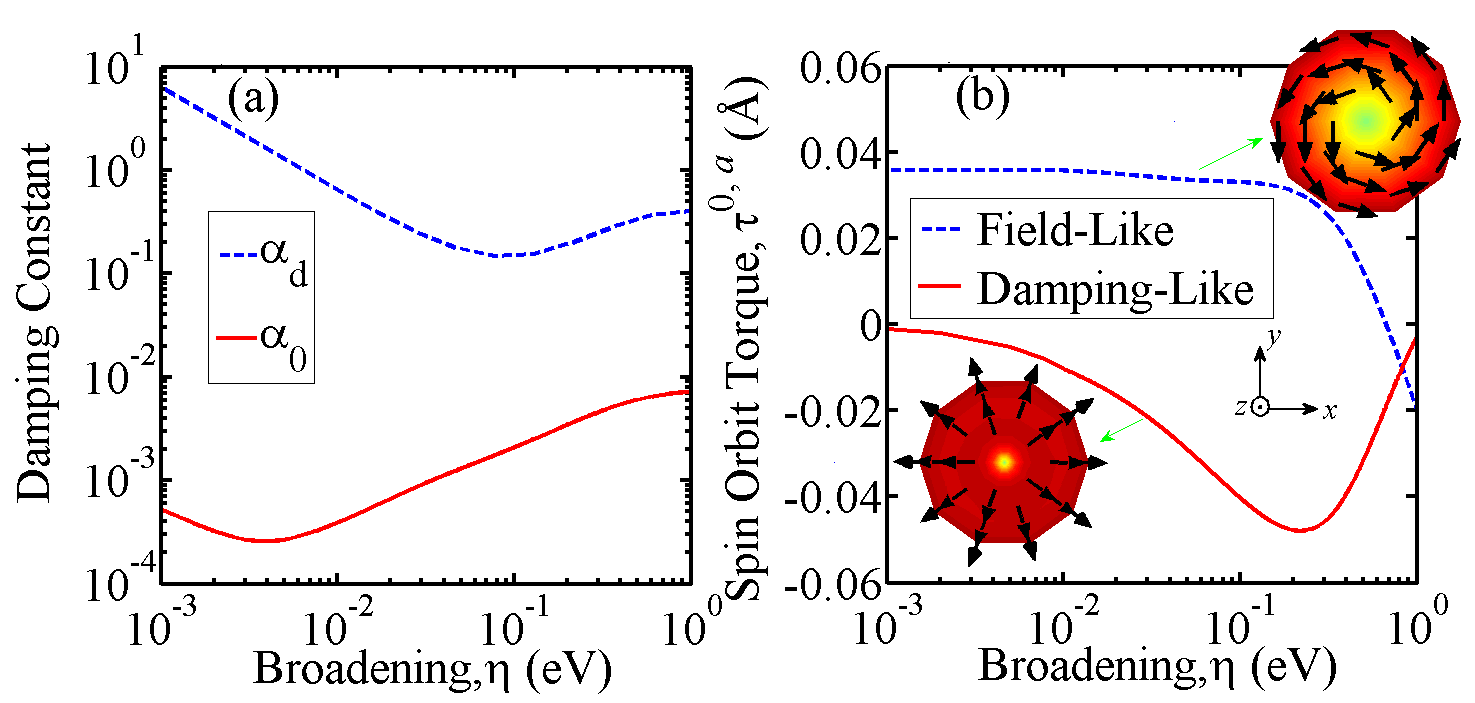}
	\caption{(Color online) (a) Sublattice Gilbert damping $\alpha_d$ (dashed blue curve) and  $\alpha_0$ (solid red curve) components for bulk cubic FeRh versus broadening parameter $\eta$. (b) Sublattice current-induced FL (dashed blue) and DL (red solid) torque coefficients for FeRh under an external electric field along the $a$-axis. The coefficients were calculated by fitting the
	vector dependence of the DL ($\propto\vec{m}_s\times(\vec{m}_s\times\vec{m}_{\bar{s}})$) and FL ($\propto\vec{m}_s\times\vec{m}_{\bar{s}}$) expressions for the symmetric and antisymmetric components in Eq. \ref{eq:Kubo-STT}, respectively. Insets display the top-view of the vector field of the FL- and DL- torques for cone angles $\le 30^o$.}
	\label{fig:fig2}
\end{figure}

  In Fig. \ref{fig:fig2}(a) we show the variation of $\alpha_d$ and $\alpha_0^{\vec{m}_s\parallel \vec {a}}$ with $\eta$ for cubic FeRh as a representative example. We find that 
  in the experimentally relevant range of $\eta$ ($\approx$ 10 - 100 meV) $\alpha_0$ is in the resistivity regime where the interband component is dominant. On the other hand,  $\alpha_d$ decreases monotonically with $\eta$, suggesting that the intraband component is dominant. 
  Unlike $\alpha_0$ which may depend on the orientation of the  N\'{e}el ordering,  $\alpha_d$ is relatively isotropic. 
  %%%%%%%%%%%%%%%%%%%%%%%%%%%%%%%%%%%%%%%%%%%%%%%%%%%%%%%%%%%%%%%%%%%%%%%%%%%%%%

	Finally, Table~\ref{table:tab1} lists the values for the current-induced FL- and DL- intersublattice torque coefficients, $\tau^{0,i}_{FL/DL}$, under an external electric field along the $i$ ($a$ or $c$) direction. The sublattice torques are determined by fixing the orientation of the $\bar{s}$ sublattice magnetization and calculating the torque for different magnetization orientations of the $s$-sublattice, using the symmetric and antisymmetric correlation expressions\cite{mahfouziPRB2018_SOT},
	
	\begin{subequations}
	\begin{align}\label{eq:Kubo-STT}
	\vec{\tau}^S_{s;i}&=\frac{\hbar}{\pi N_kM_s}\vec{m}_s\times\sum_{\vec{k}}\Tr\Big(\hat{A}_{\vec{k}}\hat{\Delta}_{\vec{k}}^s\vec{\hat{\sigma}}\hat{A}_{\vec{k}}\frac{\partial\hat{H}_{\vec{k}}}{\partial k_i}\Big),	 
	\end{align}	
	\begin{align}\label{eq:Berry-STT}
	\vec{\tau}^{AS}_{s;i}&=\frac{2}{M_sN_k}\vec{m}_s\times\sum_{nm\vec{k}}Re\left[\frac{Im((\hat{\Delta}_{\vec{k}}^s\vec{\hat{\sigma}})_{nm}(\frac{\partial\hat{H}_{\vec{k}}}{\partial k_i})_{mn})}{(\varepsilon_{n\vec{k}}-\varepsilon_{m\vec{k}}-i\eta)^2}\right]f_{n\vec{k}}. 
	\end{align}
	\end{subequations}
	Here, $f_{n\vec{k}}$ is the Fermi-Dirac distribution function and $\varepsilon_{m\vec{k}}$ are the eigenvalues of the Hamiltonian  $\hat{H}_{\vec{k}}$. Having determined the 
	torques, we fit the results to the expected $\tau_{FL}^{0,i}\vec{m}_s\times\vec{m}_{\bar{s}}$ and  $\tau_{DL}^{0,i}\vec{m}_s\times(\vec{m}_s\times\vec{m}_{\bar{s}})$ expressions and find the values for the FL and DL torque coefficients. 
	The calculations reveal that the symmetric (anti-symmetric) torque expression leads to the 
	DL (FL) component, in contrast to the SOT results in HM/FM bilayers\cite{mahfouziPRB2018_SOT}.

	Fig.~\ref{fig:fig2}(b) displays the current-induced FL and DL intersublattice torques under an external electric field along the $a$ direction for FeRh, as a representative example, versus the broadening parameter $\eta$. Note, the FL component that originates from the antisymmetric torque term [Eq.~\ref{eq:Berry-STT}] is relatively insensitive to $\eta$ (or temperature). On the other hand, the DL intersublattice torque varies almost linearly with $\eta$ (for $\eta$ <0.1 eV) and is
of extrinsic origin. Thus, in the ballistic regime where the electronic spin diffusion length is infinite, there is no current-induced transfer of angular momentum between the two sublattices, as it would violate the conservation law of total angular momentum. In the extreme opposite limit, where the spin diffusion length is much smaller than the lattice constant, each sublattice can be viewed as a magnetic lead in a spin  valve system where the intersublattice DL torque is analogous to the DL-spin transfer torque.

%%%%%%%%%%%%%%%%%%%%%%%%%%%%%%%  CONCLUSION %%%%%%%%%%%%%%%%%%%%%%%%%%%%%%%%%%%%%%%%%%%%%%%	
%	\section{Concluding remarks}\label{sec:conclusions}
	 Im summary, we have employed {\it ab-initio} based calculations to investigate the  AFMR  phenomena in
	 MnX (X=Ir,Pt,Pd,Rh)  and biaxially strained FeRh metallic AFMs in the presence or absence of an external electric field. We demonstrate that both the AFMR linewidth and effective Gilbert damping parameter can be separated into a relativistic and exchange contributions, where the former dominates at room temperature while the latter becomes significant at low temperatures. We find that both the AFMR linewidth and 
	 the intersublattice exchange interaction (and hence the AFMR frequency and N\'{e}el temperature) can be tuned by the external electric field. For example for AFM FeRh an external electric field of  1 V/$\mu m$ (current density of $\approx$ 10$^{12}$ A/m$^2$) yields an intersublattice 
	 FL torque of 3.3 meV ($\approx 0.01\lambda$)  and DL torque of 1.4 meV $\equiv$ 2.1 THz change of AFMR linewidth.

%	\begin{acknowledgments}
		
		The work is supported by NSF ERC-Translational Applications of Nanoscale Multiferroic Systems (TANMS)- Grant No. 1160504 and by NSF-Partnership in Research and Education in Materials (PREM) Grant Nos. DMR-1205734 and DMR-1828019.
%	\end{acknowledgments}
	
	\appendix

	\section{Density Functional Theory Calculations}\label{app:DFT_calc}
	We have carried out density functional theory (DFT) calculations for the MnX (X=Pt,Pd,Ir,Rh) family of metallic bulk AFMs (L1$_0$ structure) and the biaxially strain G-AFM FeRh  (bcc B2 structure) shown in Fig. 1 of the main text. 
	The DFT calculations employed the Vienna \emph{ab initio} simulation package (VASP) \cite{Kresse96a,Kresse96b}. The pseudopotential and wave functions are treated within the projector-augmented wave (PAW) method \cite{Blochl94,KressePAW}. Structural relaxations were carried using the generalized gradient approximation as parameterized by Perdew {\it{et al.}} \cite{PBE} where the largest atomic force is smaller than 0.01 eV/\AA. The plane wave cutoff energy was 500 eV and a 15 $\times$ 15 $\times$ 15 $k$ points mesh was used in the 3D Brillouin Zone (BZ) sampling for the self consistent charge relaxation. A $k$-point mesh of 8 $\times$ 8 $\times$ 8 $k$ was used to obtain the tight-binding Hamiltonian in Wannier basis set using the VASP-Wannier90 calculations \cite{Mostofi}. 
	The time-dependent electronic Hamiltonian of the system is given by,
	\begin{align}\label{eq:Hamil}
	\hat{H}_{\vec{k}}&=\hat{H}_{\vec{k}}^0\hat{1}_{2\times 2}+\sum_{s}\hat{\Delta}^s_{\vec{k}}\vec{m}_s(t)\cdot\vec{\hat{\sigma}}+\hat{H}_{SOC},
	\end{align}	
	where, $\hat{H}_{\vec{k}}^0$ is the spin-independent term of the Hamiltonian where for simplicity we have dropped the  Kronecker  matrix product symbol between matrices in the orbital and spin Hilbert spaces and $\hat{H}_{SOC}=\sum_{I,l}\xi_{I,l}\vec{\hat{L}}_{I,l}\cdot\vec{\hat{\sigma}}$ is the spin orbit term of the Hamiltonian. The  $\vec{\hat{L}}$ and $\xi_{I,l}$ are the angular momentum operator and spin-orbit coupling strength for orbital $l$ of the $I$th atom, respectively.  In the following subsections we present the details of the methods that were used to calculate the various 
	physical quantities in Table I.
	\subsection{Magneto-Crystalline Anisotropy Energy}\label{sec:MCA_calc}
\begin{figure}
		\includegraphics[scale=0.35,angle=0]{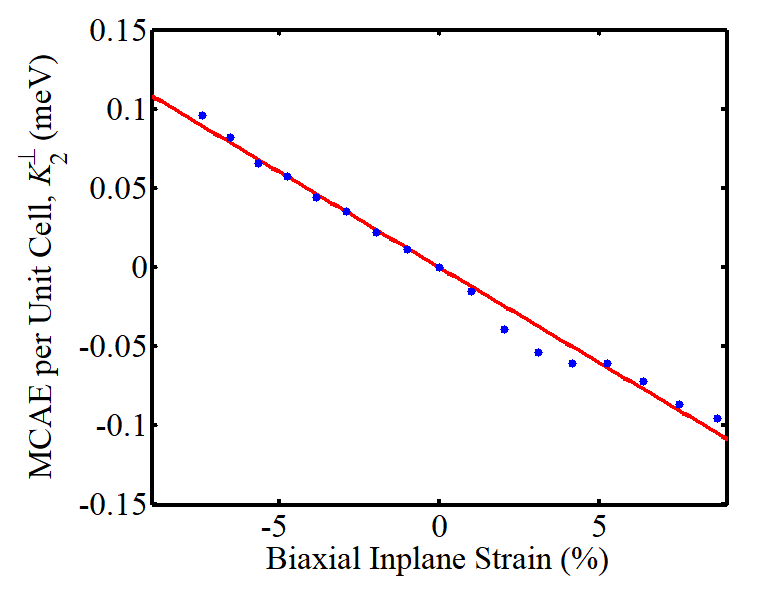}
		\caption{(Color online) Magneto Crystalline anisotropy energy versus biaxial $x=c/a-1$ strain for bulk FeRh. }
		\label{fig:figS2}
	\end{figure}

	The uniaxial magneto-crystalline anisotropy energy, $K^{\perp}_2$ was determined from the total energy difference between in-plane, $\vec{m}\parallel a$, and out of plane, $\vec{m}\parallel c$ magnetization orientations, $K^{\perp}_2=E^{tot}_{\vec{m}\parallel a}-E^{tot}_{\vec{m}\parallel c}$. For the FeRh where $c/a$= 1.0 the MCAE is small. Thus, we have calculated the variation of the MCAE with biaxial strain $x=c/a -1$ which is shown in Fig.~\ref{fig:figS2}. The MCAW can be fitted by $K_2^{\perp}=-1.2(c/a-1) meV/u.c.$, indicating that under tensile (compressive) biaxial strain the magnetization orientation is along the $a$ ($c$ axis), in agreement with previous {\it ab initio} 
	calculations. \cite{Bordel2012}
	\subsection{Calculation of Intersublattice Exchange Coupling}\label{sec:lambda_calc}
\begin{figure}
		\includegraphics[scale=0.35,angle=0]{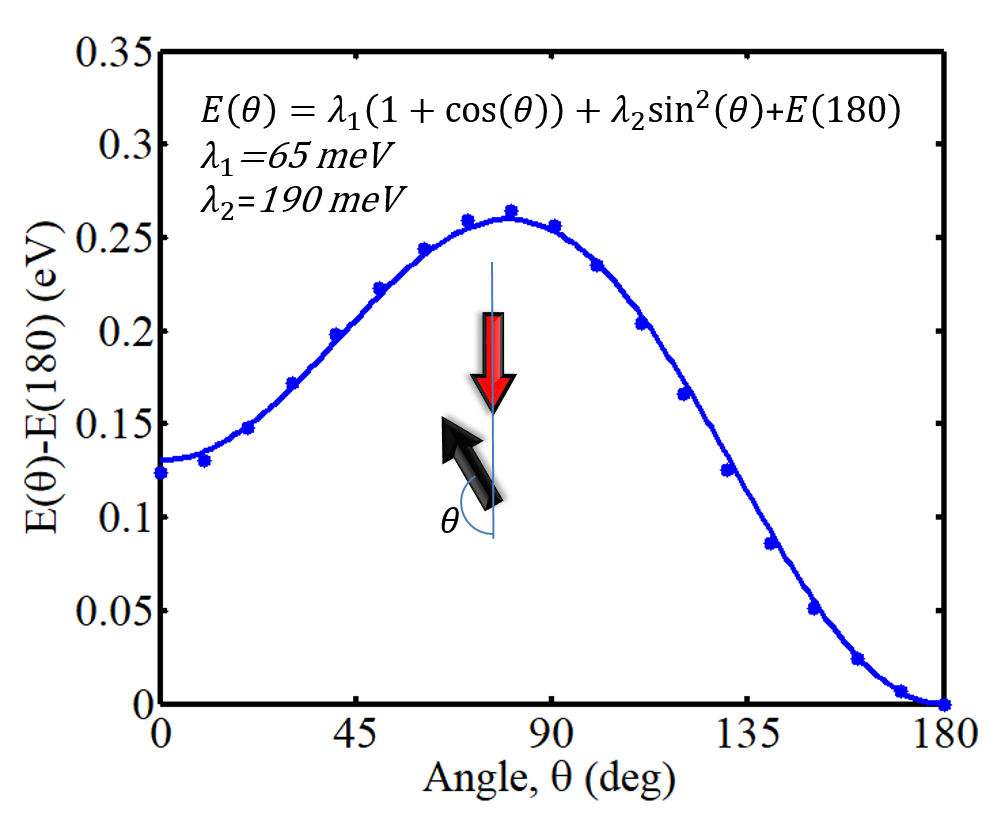}
		\caption{(Color online) Change of total energy per unit cell versus the angle between the two AFM sublattices (filled circles) for FeRh. The blue curve is the fit  to $E(\theta)-E(180)=\lambda_1(1+\cos(\theta))+\lambda_2\sin^2(\theta)$ expression.}
		\label{fig:figS3}
	\end{figure}

	The intersublattice exchange coupling is determined from total energy calculations with SOC where one varies the the angle between the sublattice magnetic moments (inset in Fig.~\ref{fig:figS3}). The total energy is calculated by imposing an orientation constrain on the magnetic moment configuration using the constrained moment method implemented in VASP where a penalty functional is added to the total energy to align the magnetic moment along a preferred direction. In Fig.~\ref{fig:figS3} we show the
	variation of $E(\theta)-E(180)$ with the angle $\theta$ (filled circles) for AFM FeRh.
	The energy difference was fitted to the  $E(\theta)-E(180^0)=\lambda_1(1+\cos(\theta))+\lambda_2\sin^2(\theta)$ expression (blue curve). The exchange coupling is in turn determined from $\lambda=\partial^2E(\theta)/\partial\theta^2|_{\theta=180^0}$, which yields $\lambda=\lambda_1+2\lambda_2=445\ meV/u.c.$. It is worth mentioning that in our calculations, $\lambda_2$, also referred to as the biquadratic exchange term\cite{Papanicolaou1988,Ivanov2003,Chubukov1990}, is significant only in the case of FeRh which undergoes an AFM to FM transition at about 350 K. The biquadratic exchange interaction provides the energy barrier for the AFM to FM transition.
	\subsection{Calculation of Conductivity}
	The longitudinal conductivity is determined from Kubo's expression, 
	\begin{align}\label{eq:Conductivity}
	\sigma_{xx}&=\frac{e^2}{\hbar}\frac{1}{\pi N_kV}\sum_{k}\Tr\Big(Im(\hat{G}_{\vec{k}}^r)\frac{\partial\hat{H}_{\vec{k}}}{\partial k_x}Im(\hat{G}_{\vec{k}}^r)\frac{\partial\hat{H}_{\vec{k}}}{\partial k_x}\Big),
	\end{align}	 
	where $V$ is the volume of the unit cell and the  resistivity is in turn given by $\rho_{xx}=1/\sigma_{xx}$. Since the relaxation time approximation is unreliable in the limit of large broadening parameter, $\eta$, we consider only the small $\eta$ limit, were the resistivity is dominated by the intraband component and is proportional to $\eta$. In this case $\rho_{xx}/\eta$ is independent of the broadening parameter that can be used to deduce an estimate of the broadening parameter by replacing the theoretical values of the resistivity with the experimental values. 
	
	\subsection{Calculation of Damping Parameters}
	
	For circular dynamics of the magnetization close to the easy axis, $\vec{m}^0_s=m^z_s\vec{e}_z$,  ($m^z_s=\pm 1$), the intersublattice damping constant tensor within torque correlation (TC) method is given by,
	\begin{align}\label{eq:Gilbert_tensor}
	\alpha_{ss'}&=\frac{\hbar}{\pi N_kM_s}\sum_{k}\Tr(Im(\hat{G}_{\vec{k}}^r)\hat{\Delta}_{\vec{k}}^s\hat{\sigma}^+Im(\hat{G}_{\vec{k}}^r)\hat{\Delta}_{\vec{k}}^{s'}\hat{\sigma}^-).
	\end{align}		
	Here, $N_k$ is the number of $k$-points in the summation, 
	$\hat{G}^r_{\vec{k}}=1/(E_F-i\eta-\hat{H}_{\vec{k}})$ is the retarded Green's function,
	$Im()$ is the imaginary part, $\hat{\sigma}^{\pm}=\hat{\sigma}_x\pm i\hat{\sigma}_y$ with $\hat{\sigma}_i$ being the Pauli matrices, 
	and  $\hat{\Delta}_{\vec{k}}^s=\frac{\hat{\Delta}_{\vec{k}}\hat{1}_s+\hat{1}_s\hat{\Delta}_{\vec{k}}}{2}$,
	is the sublattice exchange splitting  where, $\hat{1}_s$ is the diagonal matrix with identity elements for orbitals corresponding to sublattice $s$ and zero elsewhere. 
	%In order to calculate $\alpha_{0}$ using SOTC formula we replace $\hat{\Delta}_{\vec{k}}^s\hat{\sigma}^{\pm}$ in Eq.\eqref{eq:Gilbert_tensor} with $[\hat{H}_{SOC},\sigma^{\pm}]$, 
	%\begin{align}\label{eq:Gilbert_SOTC}
	%\alpha_{0}&=\frac{\hbar}{\pi N_kM}\sum_{k}\Tr(Im(\hat{G}_{\vec{k}}^r)[\hat{H}_{SOC},\sigma^{+}]Im(\hat{G}_{\vec{k}}^r)[\hat{H}_{SOC},\sigma^{-}]),
	%\end{align}	
	%where, $[,]$ represents the anti-commutation operation.

	%%%%%%%%%%%%%%%%%%%%%%%%%%%%%%%%%%%%%%%%%%%%%%%%%%%%%%%%%%%%%%%%%%%%%%%%%%%%%%%%%%%%%%%%%%%%%%%%%%%%%%%
	\section{Toy Model for Gilbert damping tensor}\label{app:B}
	It is instructive to apply the approach of the Gilbert damping constant tensor
	to a toy model and calculate the matrix elements, analytically. The simplest AFM toy model consists of a four band model Hamiltonian without SOC, $\hat{H}(\vec{k})=\varepsilon(\vec{k})\hat{1}+\Delta\hat{\tau}_z\vec{m}\cdot\vec{\hat{\sigma}}+T\hat{\tau}_x$, where, $\hat{\sigma}_i$s are the Pauli matrices, $\hat{\tau}_i$s are Pauli matrices in sublattice space and $T$ is the intersublattice hopping parameter. 
	
	For the intraband component of the intrasublattice damping parameter tensor elements we obtain, $\alpha_{ss}=\frac{1}{4MN_k}\sum_{n\vec{k}}\frac{T^2\Delta^2}{\Delta^2+T^2}\delta(E_F-\varepsilon_n)^2$, where $\varepsilon_{1,2}(\vec{k})=\varepsilon(\vec{k})\pm\sqrt{T^2+\Delta^2}$. Within the relaxation time ($\tau_{el}$) approximation and introducing the parameter $\eta=\hbar/2\tau_{el}$ to broaden the Dirac delta function we find, $\alpha_{ss}\approx\frac{1}{8M\eta\pi}\frac{T^2\Delta^2}{\Delta^2+T^2}g_0$, where $g_0=\frac{2}{N_k}\sum_{n\vec{k}}\delta(E_F-\varepsilon_n)$ is the density of states per unit cell at the Fermi energy.  Similarly, for the intersublattice elements we obtain, $\alpha_{s\bar{s}}=\alpha_{ss}$, where, as expected due to the absence of the SOC  $\alpha_{0}=0$.  This suggests that while the microscopic origin of the intrinsic damping, $\alpha_{0}$, is rooted in the transfer of the angular momentum from local spin moments to the crystal mediated by the SOC, the individual sublattice Gilbert damping tensor elements, $\alpha_{ss'}$ are governed by the hopping strength of the electrons between different sublattices.

	\section{Derivation of AFMR Frequency and Linewidth}\label{app:C}	
	Since, experimental measurements of the magnetic resonance phenomena is often performed by sweeping the amplitude of the time independent external magnetic field and fixed frequency for the microwave, we define $\beta_{ss'}=i\alpha_{ss'}\omega$, where $\omega$ is the microwave frequency. Eq.1 in the main text for an AFM with, $m_{1}^z=-m_{2}^z$, $M=M_{s}$ and $K_{2;s}^{x,y}=K_2^{x,y}$, can be rewritten as,
	\begin{subequations}
		\begin{align}
		&i\omega m_s^x -\gamma\vec{\tau}_{DL}\cdot\vec{E}_{ext}(m_{s}^x+m_{\bar{s}}^x)=\sum_{s'}\Omega^y_{ss'}m_{s'}^y\\
		&i\omega m_s^y -\gamma\vec{\tau}_{DL}\cdot\vec{E}_{ext}(m_{s}^y+m_{\bar{s}}^y)=-\sum_{s'}\Omega^x_{ss'}m_{s'}^x,
		\end{align}
	\end{subequations}
	where,
	\begin{align}
	\hat{\Omega}^j&=\gamma B^z_{ext}\hat{1}+\frac{\gamma}{M}\left((\mathcal{K}_{j}+\beta_d+\lambda)\hat{\sigma}_z+i(\lambda+\beta_{\bar{d}})\hat{\sigma}_y\right).
	\end{align}
	Here, we assumed $K^{z}_2=0$, $\beta_{d}=\beta_{ss}$ and  $\beta_{\bar{d}}=\beta_{s\bar{s}}$. The eigen-frequencies of the system are given by
	\begin{widetext}
	\begin{align}\label{eq:gen_AFMR_exp}
	&(\omega/\gamma-i\vec{\tau}_{DL}\cdot\vec{E}_{ext})^2 =\Omega_{\parallel}^2 + (B^z_{ext})^2 - \frac{\mathcal{K}_{\perp}^2}{M^2} -(\vec{\tau}_{DL}\cdot\vec{E}_{ext})^2\pm 2\omega_0, \nonumber
	\end{align}
	where,
	\begin{align}
	&\Omega^2_{\parallel}=\frac{(\mathcal{K}_{\parallel}+\beta_{d}-\beta_{\bar{d}})(\mathcal{K}_{\parallel}+\beta_{d}+\beta_{\bar{d}}+2\lambda)}{M^2},\\
	&\omega_0^2=( B^z_{ext})^2\left(\Omega^2_{\parallel} - (\vec{\tau}_{DL}\cdot\vec{E}_{ext})^2\right) + \frac{\mathcal{K}_{\perp}^2(\lambda+\beta_{\bar{d}})^2}{M^4},
	\end{align}
	and, we define, $\mathcal{K}_{\parallel}=(\mathcal{K}_{x}+\mathcal{K}_{y})/2$ and $\mathcal{K}_{\perp}=(\mathcal{K}_{x}-\mathcal{K}_{y})/2$. 
	In the absence of an external magnetic field and in linear response regime to the external electric field we obtain,
	\begin{align}
	&(\omega/\gamma-i\vec{\tau}_{DL}\cdot\vec{E}_{ext})^2=\frac{(\mathcal{K}_{\parallel}\mp\mathcal{K}_{\perp}+\beta_{d}-\beta_{\bar{d}})(2\lambda+\mathcal{K}_{\parallel}\pm\mathcal{K}_{\perp}+\beta_{d}+\beta_{\bar{d}})}{M^2}\nonumber
	\end{align}
	\end{widetext}

	%BibTeX
	%Windows:
	%\bibliographystyle{D:/PHYSICS/TEX/BIBTEX/prsty}
	%\bibliography{D:/PHYSICS/TEX/BIBTEX/qttg}
	
	%Linux:
	%\bibliographystyle{apsrev}
	%\bibliography{$HOME/TEX/BIBTEX/qttg}

\begin{thebibliography}{10}
		
		\bibitem{Slonczewski1996}  J. C. Slonczewski, Current-driven excitation of magnetic multilayers, J. Magn. Magn. Mater. {\bf 159},  L1-L7  (1996).
		
		\bibitem{Berger1996}  L. Berger, Emission of spin waves by a magnetic multilayer traversed by a current, Phys. Rev. B {\bf 54},  9353  (1996).
		
		\bibitem{Manchon2008}				
		A. Manchon and S. Zhang, Theory of nonequilibrium intrinsic spin torque in a single nanomagnet,				
		Phys. Rev. B {\bf 78}, 212405, (2008).
		%%%%%%%%%%%%%%%%%%%%%%%%%%%%%%%%%%%%%%%%%%%%%%%%%%%%%%%%%%%%%%%%%%%%%%%%%%%%%%%%%%%%%%%%%
		\bibitem{Miron2010}
		Ioan Mihai Miron,	Gilles Gaudin,	St\'{e}phane Auffret,	Bernard Rodmacq,	Alain Schuhl, Stefania Pizzini,	Jan Vogel and Pietro Gambardella, Current-driven spin torque induced by the Rashba effect in a ferromagnetic metal layer, Nature Materials {\bf 9}, 230-234 (2010).
		
		\bibitem{Miron2011}
		Ioan Mihai Miron,	Kevin Garello,	Gilles Gaudin,	Pierre-Jean Zermatten,	Marius V. Costache, St\'{e}phane Auffret,	S\'{e}bastien Bandiera,	Bernard Rodmacq, Alain Schuhl and Pietro Gambardella, Perpendicular switching of a single ferromagnetic layer induced by in-plane current injection, Nature {\bf 476}, 189-193 (2011).
		%%%%%%%%%%%%%%%%%%%%%%%%%%%%%%%%%%%%%%%%%%%%%%%%%%%%%%%%%%%%%%%%%%%%%%%%%%%%%%%%%%%%
		\bibitem{Liu2012}
		Luqiao Liu, O. J. Lee, T. J. Gudmundsen, D. C. Ralph, and R. A. Buhrman, Current-Induced Switching of Perpendicularly Magnetized Magnetic Layers
		Using Spin Torque from the Spin Hall Effect, Phys. Rev. Lett. {\bf 109}, 096602 (2012).
		
		\bibitem{Baltz2018}
		Antiferromagnetic spintronics,
		V. Baltz, A. Manchon, M. Tsoi, T. Moriyama, T. Ono, and Y. Tserkovnyak
		Rev. Mod. Phys. 90, 015005 (2018).
		
		\bibitem{Gomonay2014}
		E. V. Gomonay,  and V. M. Loktev, Spintronics of antiferromagnetic systems (review article). Low Temp. Phys. 40, 17-35 (2014).
		
		\bibitem{Kirilyuk2010}
		A. Kirilyuk, A. V. Kimel, and T. Rasing, Ultrafast optical manipulation of magnetic order, Rev. Mod. Phys. {\bf 82}, 2731 (2010).

		\bibitem{Wienholdt2012}
		S. Wienholdt, D. Hinzke, and U. Nowak,
		THz Switching of Antiferromagnets and Ferrimagnets, 
		Phys. Rev. Lett. {\bf 108}, 247207 (2012).

				
		\bibitem{Wadley2016}
		P. Wadley, B. Howells, J. Zelezny, C. Andrews, V. Hills, R. P. Campion, V. Novak, K. Olejnik, F. Maccherozzi, S. S. Dhesi, S. Y. Martin, T. Wagner, J. Wunderlich, F. Freimuth, Y. Mokrousov, J. Kunes, J. S. Chauhan, M. J. Grzybowski, A. W. Rushforth, K. W. Edmonds, B. L. Gallagher, and T. Jungwirth, Electrical switching of	an antiferromagnet, Science {\bf 351}, 587 (2016).
				
		\bibitem{Bhattacharjee2018}
		N. Bhattacharjee, A. A. Sapozhnik, S. Yu. Bodnar, V. Yu. Grigorev, S. Y. Agustsson, J. Cao, D. Dominko,
		M. Obergfell, O. Gomonay, J. Sinova, M. Kläui, H.-J. Elmers, M. Jourdan, and J. Demsar, N\'{e}el Spin-Orbit Torque Driven Antiferromagnetic Resonance in Mn2Au
		Probed by Time-Domain THz Spectroscopy, Phys. Rev. Lett. {\bf 120}, 237201 (2018).
				
		\bibitem{Kambersky2007}
		V. Kambersky, Spin-orbital Gilbert damping in common magnetic metals,
		Phys. Rev. B {\bf 76}, 134416 (2007).
				
		\bibitem{mahfouziPRB2017_GD}
		F. Mahfouzi, JJinwoong Kim, and N. Kioussis,
		Intrinsic damping phenomena from quantum to classical magnets: An ab initio study of Gilbert damping in a Pt/Co bilayer,
		Phys. Rev. B {\bf 96}, 214421 (2017).
		
		\bibitem{Dykanov1971}
		M.I. Dyakonov and V.I. Perel, Current-induced spin orientation of electrons in semiconductors,
		Phys. Lett. A {\bf 35}, 459 (1971).
		
		\bibitem{Sinova2015}
		Jairo Sinova, Sergio O. Valenzuela, J. Wunderlich, C. H. Back, and T. Jungwirth,		
		Spin Hall effects,
		Rev. Mod. Phys. {\bf 87}, 1213 (2015).
		
		\bibitem{Ralph2008}
		D. C. Ralph, and M. D. Stiles, Spin transfer torques,
		J. Magn. Magn. Mater. {\bf 320}, 1190 (2008).
		
		\bibitem{Nunez2006}
		A. S. Nunez, R. A. Duine, Paul Haney, and A. H. MacDonald,
		Theory of spin torques and giant magnetoresistance in antiferromagnetic metals
		Phys. Rev. B {\bf 73}, 214426 (2006).
		
		\bibitem{Gulyaev2014}
		Y. V. Gulyaev, P. E. Zilberman, G. M. Mikhailov and S. G. Chigarev, 
		Generation of terahertz waves by a current in magnetic junctions,
		JETP Lett. 98, 742-752 (2014).
		
		\bibitem{Cheng2016}
		R. Cheng, Di Xiao, and A. Brataas,
		Terahertz Antiferromagnetic Spin Hall Nano-Oscillator,
		Phys. Rev. Lett. {\bf 116}, 207603 (2016).
		
		\bibitem{Gulbrandsen2018} 		
		S. A. Gulbrandsen and A. Brataas, 
		Spin transfer and spin pumping in disordered normal metal-antiferromagnetic insulator systems,
		Phys. Rev. B {\bf 97}, 054409 (2018).
		
		\bibitem{Khymyn2017}
		R. Khymyn, I. Lisenkov, V. Tiberkevich, B. A. Ivanov and A. Slavin, Antiferromagnetic THz-frequency Josephson-like Oscillator Driven by Spin Current,
		Scientific Reports {\bf 7}, 43705 (2017).
		
		\bibitem{Kriegner2016} 
		D. Kriegner, K. V\'{y}born\'{y}, K. Olejnik, H. Reichlov\'{a}, V. Nov\'{a}k, X. Marti, J. Gazquez, V. Saidl, P. Nemec, V. V. Volobuev, G. Springholz, V. Hol\'{y} and T. Jungwirth,
		Multiple-stable anisotropic magnetoresistance memory in antiferromagnetic MnTe. Nat. Commun. {\bf 7}, 11623
		(2016).
		
		\bibitem{Chen2018} 
		X. Z. Chen, R. Zarzuela, J. Zhang, C. Song, X. F. Zhou, G. Y. Shi, F. Li,
		H. A. Zhou, W. J. Jiang, F. Pan, and Y. Tserkovnyak, Antidamping-Torque-Induced Switching in Biaxial Antiferromagnetic Insulators,
		Phys. Rel. Lett. {\bf 120} 207204 (2018).
		
		\bibitem{Moriyama12018} 
		T. Moriyama, K. Oda, T. Ohkochi, M. Kimata and T. Ono, Spin torque control of antiferromagnetic moments in NiO,
		Scientific Reports {\bf 8} 14167 (2018).
		
		\bibitem{Zelezny2014}
		J. Zelezny, H. Gao, K. Vyborny, J. Zemen, J. Masek, A. Manchon, J. Wunderlich,
		Jairo Sinova, and T. Jungwirth,
		Relativistic N\'{e}el-Order Fields Induced by Electrical Current in Antiferromagnets,		
		Phys. Rev. Lett. {\bf 113}, 157201 (2014).
		
		\bibitem{Zelezny2017}
		J. Zelezny, H. Gao, A. Manchon, F. Freimuth, Y. Mokrousov, J. Zemen, J. Masek,
		J. Sinova, and T. Jungwirth,
		Spin-orbit torques in locally and globally noncentrosymmetric crystals:
		Antiferromagnets and ferromagnets
		Phys. Rev. B {\bf 95}, 014403 (2017).
		
		\bibitem{Kittel1951}
		F. Keffer and C. Kittel, Theory of Antiferromagnetic Resonance,
		Phys. Rev. {\bf 85}, 329 (1952).		
		
		\bibitem{Mentink2012}
		J. H. Mentink, J. Hellsvik, D. V. Afanasiev, B. A. Ivanov, A. Kirilyuk, A. V. Kimel, O. Eriksson, M. I. Katsnelson, and Th. Rasing,
		Ultrafast Spin Dynamics in Multisublattice Magnets,
		Phys. Rev. Lett. {\bf 108}, 057202 (2012).
		
		
		\bibitem{Liu2017}
		Q. Liu, H. Y. Yuan, K. Xia, and Z. Yuan,
		Mode-dependent damping in metallic antiferromagnets due to intersublattice spin pumping,
		Phys. Rev. Materials {\bf 1}, 061401(R) (2017).
		
		\bibitem{Checinski2017}
		J. Ch\k{e}ci\'{n}ski, M. Frankowski, and T. Stobiecki, Antiferromagnetic nano-oscillator in external magnetic fields,
		Phys. Rev. B {\bf 96}, 174438, (2017).
		
		
		
		\bibitem{Mankovsky2017} S. Mankovsky, S. Polesya, K. Chadova, H. Ebert, J. B. Staunton, T. Gruenbaum, M. A. W. Schoen, C. H. Back, X. Z. Chen, and C. Song, Temperature-dependent transport properties of FeRh,
		Phys. Rev. B {\bf 95},  155139 (2017).
		
		\bibitem{KOUVEL1963} J. S. Kouvel, C. C. Hartelius, and L. M. Osika, Magnetic Properties and Crystal-Structure Transformation of the Ordered Alloy (MnRh),
		Journal of Appl. Phys. {\bf 34}, 4 (1963).
		
		\bibitem{Zhang2014} W. Zhang, M. B. Jungfleisch, W. Jiang, J. E. Pearson, A. Hoffmann, F. Freimuth, and Y. Mokrousov, Spin Hall Effects in Metallic Antiferromagnets,
		Phys. Rev. Lett. {\bf 113},  196602 (2014).
		
		\bibitem{Ou2016} Y. Ou, S. Shi, D. C. Ralph, and R. A. Buhrman, Strong spin Hall effect in the antiferromagnet PtMn,
		Phys. Rev. B {\bf 93},  220405(R) (2016).
		
		\bibitem{Kim2016} 
		D. J. Kim,  K. D. Lee,  S. Surabhi,  S. G. Yoo, J. R.Jeong, B. G. Park, Utilization of the Antiferromagnetic IrMn Electrode in Spin Thermoelectric Devices and Their Beneficial Hybrid for Thermopiles,
		Adv. Funct. Mater. \textbf{26}, 30 (2016).
		
		\bibitem{Moriyama2014} T. Moriyama, M. Nagata, K. Tanaka, K.-j. Kim, H. Almasi, W. Wang, and T. Ono, Spin-transfer-torque
		through antiferromagnetic IrMn,
		arXiv:1411.4100, arXiv (2014).
		
		\bibitem{Tshitoyan2015}V. Tshitoyan, C. Ciccarelli, A. P. Mihai, M. Ali, A. C. Irvine, T. A. Moore, T. Jungwirth, and A. J. Ferguson, Electrical manipulation of ferromagnetic NiFe by antiferromagnetic IrMn,
		Phys. Rev. B {\bf 92}, 214406 (2015).
		
		
		\bibitem{Bordel2012} 
		C. Bordel, J. Juraszek, D. W. Cooke, C. Baldasseroni, S. Mankovsky, J. Minar, H. Ebert, S. Moyerman,
		E. E. Fullerton, and F. Hellman, Fe Spin Reorientation across the Metamagnetic Transition in Strained FeRh Thin Films,
		Phys. Rev. Lett. {\bf 109}, 117201 (2012).
		
		
		\bibitem{Umetsu2006} 
		R. Y. Umetsu, A. Sakuma, and K. Fukamichi, Magnetic anisotropy energy of antiferromagnetic -type equiatomic Mn alloys,
		Appl. Phys. Lett. {\bf 89}, 052504 (2006).		
		
		\bibitem{Shick2010} 
		A. B. Shick, S. Khmelevskyi, O. N. Mryasov, J. Wunderlich, and T. Jungwirth, Spin-orbit coupling induced anisotropy effects in bimetallic antiferromagnets:
		A route towards antiferromagnetic spintronics, Phys. Rev. B {\bf 81}, 212409 (2010).
		
		\bibitem{Chang2018} 
		P.-H. Chang, I. A. Zhuravlev, and K. D. Belashchenko,
		Origin of spin reorientation transitions in antiferromagnetic MnPt-based alloys,
		Phys. Rev. Materials {\bf 2}, 044407 (2018).
		
		
		\bibitem{mahfouziPRB2018_SOT}
		F. Mahfouzi and N. Kioussis, First-principles study of the angular dependence of the spin-orbit torque in Pt/Co and Pd/Co bilayers,		
		Phys. Rev. B. {\bf 97}, 224426 (2018).
		% % % % % % % % % % % % % % % % % % SM % % % % % % %
		
		
		%%%%%%%%%%%%%%%%%%%%%%%%%%%%%%%%%%%%%%%%%   DFT  %%%%%%%%%%%%%%%%%%%%%%%%%%%%%%%%%%%%%%%%%%%%%%
		% VASP
		\bibitem{Kresse96a} G. Kresse and J. Furthm\"uller, Efficient iterative schemes for ab initio total-energy calculations using a plane-wave basis set,
		Phys. Rev. B \textbf{54}, 11169 (1996).
		% VASP
		\bibitem{Kresse96b}  G. Kresse and J. Furthm\"uller, Efficiency of ab-initio total energy calculations for metals and semiconductors using a plane-wave basis set,
		Comput. Mater. Sci. \textbf{6}, 15 (1996).
		% PAW
		\bibitem{Blochl94} P. E. Bl\"ochl, Projector augmented-wave method,
		Phys. Rev. B \textbf{50}, 17953 (1994).
		% VASP-PAW
		\bibitem{KressePAW} G. Kresse and D. Joubert, From ultrasoft pseudopotentials to the projector augmented-wave method,
		Phys. Rev. B \textbf{59}, 1758 (1999).
		% PBE
		\bibitem{PBE} J. P. Perdew, K. Burke, and M. Ernzerhof, Generalized Gradient Approximation Made Simple,
		Phys. Rev. Lett. \textbf{77}, 3865 (1996).
		% Wannier90
		\bibitem{Mostofi} A. A. Mostofi, J. R. Yates, G. Pizzi, Y.-S. Lee, I. Souza, D. Vanderbilt, and N. Marzari, An updated version of wannier90: A tool for obtaining maximally-localised Wannier functions,
		Comput. Phys. Commun. \textbf{185}, 2309 (2014).		
		%%%%%%%%%%%%%%%%%%%%%%%%%%%%%%%%%%%%%%%%%%%%%%%%%%%%%%%%%%%%%%%%%%%%%%%%%%%%%%%%%%%	
		\bibitem{Papanicolaou1988}
		N. Papanicolaou Unusual phases in quantum spin-1 systems,
		Nuclear Physics {\bf B305}, 367-395 (1988).
		\bibitem{Ivanov2003}
		B. A. Ivanov and A. K. Kolezhuk,
		Effective field theory for the S=1 quantum nematic,
		Phys. Rev. B {\bf 68}, 052401 (2003).
		
		\bibitem{Chubukov1990}
		A. V. Chubukov,
		Unusual states in the Heisenberg model with competing interactions,
		J. Phys.- Cond. Matter. {\bf 2}, 455 (1990).
		
		
		
		
	\end{thebibliography}
	
\end{document}